\documentstyle[11pt]{article}

\begin{document}

\begin{center}
\Large\bf
The Solar Test of the Equivalence Principle\\
\end{center}

\vskip 26pt
\begin{center}
{\sl John D. Anderson,
\footnote{ E-mail: jda@zeus.jpl.nasa.gov}
Mark Gross,
\footnote{Department of Physics and Astronomy,
California State University,
Long

\hskip 7pt Beach CA 90840.
 E-mail: mgross@csulb.edu }
Kenneth L. Nordtvedt
\footnote{Northwest Analysis, 118 Sourdough Ridge Road,
Bozeman MT 59715}
and Slava G. Turyshev
\footnote{On leave from Bogolyubov Institute for
Theoretical Microphysics, Moscow State University,

\hskip 7pt Moscow, 119899 Russia. E-mail: sgt@zeus.jpl.nasa.gov}}
 \vskip 0.7cm
\centerline{\sl Jet Propulsion Laboratory MS 301-230,}
\centerline{\sl California Institute of Technology}
\centerline{\sl 4800 Oak Grove Drive - Pasadena,
CA 91109 - USA}
\end{center}
\vskip 10mm

\begin{abstract}
The Earth, Mars, Sun, Jupiter system   allows for a sensitive
test of the strong equivalence principle (SEP) which is
qualitatively different from that provided by Lunar
Laser Ranging.
 Using analytic and numerical methods we demonstrate
 that  Earth-Mars ranging can provide a
useful estimate of the SEP  parameter $\eta$.
Two estimates of the predicted accuracy   are
derived and quoted, one based on conventional   covariance
 analysis, and another (called ``modified worst case''
analysis) which
assumes that systematic errors dominate the experiment.
If future Mars missions provide ranging measurements
with an accuracy of $\sigma$ meters, after ten years of
ranging the expected  accuracy for the SEP parameter
$\eta$  will be  of order $(1-12)\times 10^{-4}\sigma $.
 These ranging measurements will also  provide the most
accurate determination of the mass of Jupiter,
independent of the  SEP effect test.

Subject headings: celestial mechanics, stellar dynamics --
gravitation -- Earth -- planets and satellites: Mars --
dark matter
\end{abstract}
\vskip 10pt

\section{Introduction}

The  question, posed long ago by Newton, of
 the relation between the gravitational and inertial
masses of the same body continues to be the subject of
theoretical and
experimental investigations. This question   arises in
most any theory of gravitation. The
equality of inertial and passive gravitational
 masses, often stated as  the weak equivalence
principle (WEP), implies that different neutral massive
test bodies will have the same acceleration of free fall
$\vec{A}_0$ in an external gravitational field, and
therefore in  freely falling
inertial frames the external gravitational
field appears only in the form of a tidal
interaction (Singe 1960).
 Up to these tidal corrections,
freely falling   bodies behave as if
external gravity were absent (Bertotti \& Grishchuk 1990).
 In the
construction of the general theory of relativity
Einstein went further,
postulating that not only mechanical
laws of motion, but all non-gravitational laws should behave
in freely falling frames as if
gravity were absent. If local gravitational physics
is also  independent of the more extended
gravitational environment,
we have what is known as the Strong Equivalence Principle (SEP).

 Various experiments have  been  performed to measure
 the ratios of gravitational to inertial masses of   bodies.
 Experiments on bodies of laboratory dimensions verify
the WEP  to a fractional precision
$\delta A_0/A_0 =  \delta {({m_g /m})} \sim 10^{-11}$ by
(Roll, Krotkov \& Dicke  1964) and more recently to a precision
 $\delta A_0/A_0 \sim 10^{-12}$ by (Braginsky \& Panov
1972; Adelberger et al. 1994).
The accuracy of these
experiments is sufficiently high to confirm equal  strong,
weak, and electromagnetic interaction contributions to both
the passive gravitational and inertial masses of the
laboratory bodies. This
impressive evidence for  laboratory size bodies does not,
however, carry over to celestial body scales.

 Laboratory size bodies used in the
experiments cited above possess a negligible
fraction of gravitational self-energy and therefore such experiments
indicate nothing about the
equality  of gravitational self-energy contributions to the
inertial and passive gravitational masses
of the bodies  (Nordtvedt 1968a). Interesting  results for
 celestial bodies
are obtained if one includes terms of fractional order
$(\Omega_B / mc^{2})$,
where $m$ is the mass of a body $B$ and $\Omega_B$ is its
gravitational binding or self-energy:
$$ \left({\Omega \over m c^{2}}\right)_B =
 -\hskip2pt{G\over 2m c^{2}}\hskip 1mm{ \int _{V_B} }
d^{3} x d^{3} y
\hskip 2mm {\rho_B(x) \rho_B(y)\over |x - y|}
\hskip5pt.  \eqno(1) $$

\noindent This ratio is typically $\sim 10^{-25}$ for bodies
 of laboratory sizes,
so experimental
accuracy of  a part in $10^{12}$ sheds no light on how
  gravitational
self-energy contributes to the inertial and
gravitational masses of  bodies.

To test the  SEP one must utilize  planetary-sized
extended  bodies  in  which case the ratio (1) is
considerably higher.
 Numerically evaluation of the integral
of expression (1) for the standard solar model
 (Ulrich  1982) obtains
$$\left({\Omega \over mc^{2}}\right)_S \approx -3.52
\hskip1pt \cdot 10^{-6}   \eqno(2a)$$

\noindent and the analogous value has
been obtained for the Earth (Allen 1985):
$$\left({\Omega \over mc^{2} }\right)_E
\approx -4.6 \cdot10^{-10}, \eqno(2b)$$

The development of the parameterized post-Newtonian
(PPN) formalism
( Nordtvedt 1968b; Will  1971; Will \&
Nordtvedt 1972), allows one to
describe within the common framework the motion of
celestial bodies
in  external gravitational fields  within
a wide class of metric theories of gravity.  Within the
accuracy of modern experimental techniques, the PPN formalism
 becomes useful framework for testing the SEP for extended
bodies. In that formalism, the ratio of
passive gravitational to inertial mass is given by
(Nordtvedt 1968a,b)
$$ {m_g \over m_i} = \hskip2pt 1 \hskip2pt +
\hskip2pt \eta \cdot {\Omega \over mc^{2}} \hskip2pt ,
\hskip10pt  \eqno(3) $$

\noindent in which the SEP violation  is quantified by
the parameter $\eta$.
In fully-conservative, Lorentz-invariant theories of gravity
the SEP parameter is related to the PPN parameters by
$$ \eta = 4 \beta - \gamma - 3  \hskip10pt  \eqno(4) $$

\noindent and is more generally related to the complete set of PPN
parameters through the relation
$$ \eta = 4 \beta - \gamma - 3 -
{10 \over 3}\hskip 1mm\xi - \alpha_1 +
{1\over 3}\hskip 1mm (2 \alpha _2 -
2 \zeta_1 - \zeta_2). \hskip10pt
\eqno(5) $$

A difference between gravitational and inertial masses
produces observable  perturbations in the
motion of celestial bodies in the Solar System. By analyzing the
effect of a non-zero $\eta$
on the dynamics of the Earth-Moon system moving in the
gravitational field of the Sun, Nordtvedt (1968c) found  a
polarization of the Moon's orbit in the direction of the Sun
with amplitude $\delta r \sim \eta C_0$, where $C_0$ is a
constant of order $13m$.
We call this effect, generalized to all similar
three body situations,
 the ``SEP polarization effect".
The most accurate test of this effect is presently
provided by Lunar Laser Ranging (LLR)
(Williams  1976; Shapiro et al. 1976; Dickey  et al. 1989),
and in the most  recent  results (Dickey  et al. 1994;
Williams   et al. 1995)  the
parameter $\eta$ was determined to be
\vskip -8pt
$$\eta = -0.0005 \pm 0.0011 \hskip2pt.   \eqno(6)$$

\noindent  Other tests of SEP violation have been discussed.
 An experiment employing existing  binary pulsar data
has been proposed by Damour and Sch$\ddot{a}$fer (1991).
A search for the SEP polarization effect in the motion of the
Trojan asteroids was  suggested in (Nordtvedt 1968a)
and carried out by
(Orelana \& Vucetich 1992).
Also results are available from numerical experiments  with
combined processing of  LLR, spacecraft tracking,
planetary radar and Very Long Baseline Interferometer (VLBI) data
(Chandler  et al. 1994).

It has been observed previously that a measurement of the Sun's
gravitational to inertial mass ratio can be
performed using the Sun-Jupiter-Mars or Sun-Jupiter-Earth system
(Nordtvedt 1970; Shapiro  et al.  1976).
This is the first paper from a planed series addressing the
above problem. The question we would like to answear first is how
accurately can we do this ranging experiment?
We emphasize that the Sun-Mars-Earth-Jupiter system, though
governed basically by the same equations of motion as
Sun-Earth-Moon system, is significantly different physically.
For a given value of SEP parameter $\eta$ the polarization
effects on the Earth and Mars orbits are
almost two orders of magnitude larger than on the lunar orbit.
 In this work we examine the SEP effect on the
Earth-Mars range, which has been measured as part of the
Mariner 9 and Viking missions.  Moreover, future Mars missions,
now being planned as joint U.S.-Russian endeavours, should
yield additional ranging data.

 The dynamics of the four-body Sun-Mars-Earth-Jupiter
system in the Solar system barycentric inertial frame
were considered.
The quasi-Newtonian acceleration of the Earth $(E)$ with respect
to the Sun $(S)$ is straightforwardly calculated to be:
$$ \vec{A}_E - \vec{A}_S =
- \mu^*_{SE}\cdot {  \vec{R}_{SE} \over R_{SE}^3} \hskip5pt +
\hskip5pt
 \Bigl ({m_g \over m_i} \Bigl )_{\hskip-2pt E}
\sum_{k=M,J} \mu_k \hskip5pt \Bigl [
{ \vec{R}_{kS}\over R_{kS}^3} - {\vec{R}_{kE} \over R_{kE}^3}
\Bigl ] \hskip5pt + $$

$$ + \hskip5pt \eta \Bigl [
\Bigl ( {\Omega \over mc^{2}} \Bigl )_{\hskip-2pt S} -
\Bigl ({\Omega \over mc^{2}} \Bigl )_{\hskip-2pt E}
{\Bigl ]} \sum_{k = M,J} \mu_k { \vec{R}_{kS} \over R_{kS}^3}   =
\hskip10pt \vec{A}_{N} + \vec{A}_{tid}  +
\vec{A}_{\eta}\hskip5pt \eqno(7) $$

\noindent where  $\mu^*_{SE} \equiv \mu_S + \mu_E + \eta \Bigl[\mu_S
\Bigl ( {\Omega \over mc^{2}} \Bigl )_{\hskip-2pt E} +
 \mu_E \Bigl ({\Omega \over mc^2} \Bigl )_{\hskip-2pt S}
\Bigl ]$ and
$\mu_k \equiv G m_k$.
The subscripts $(M)$ and $(J)$ indicate
Mars and Jupiter, respectively. Also
$\vec{R}_{BC}=\vec{R}_{C}-\vec{R}_{B}$ is the vector
from body $B$ to body $C$ and $\vec{A}_N$ is the
Newtonian acceleration term.
$\vec{A}_{\eta}$ is the SEP acceleration term, which is of order
$1/c^2$. While it is not the only term of
that order,  the other post-Newtonian $1/c^2$ terms
(suppressed in eq.(7)) do not
affect the determination of $\eta$
until the second post-Newtonian order ($\sim 1/c^4$). Finally
$\vec{A}_{tid}$ is the Newtonian tidal acceleration term.
Note, that $A_{\eta}/A_N \sim \eta \cdot10^{-10}$ \hskip2pt and
\hskip2pt $A_{tid}/A_N\sim 7 \cdot 10^{-6}$.
Given that level of accuracy, we ignore the mutual
attraction of the two planets, Earth and Mars.
The SEP acceleration is treated as a perturbation on the
restricted three-body problem, and   the SEP
 effect is evaluated as an alteration of the planetary Keplerian orbit.

Using expression (3) and noticing that $\mu_M/R_{SM}  \ll \mu_J/R_{SJ}$,
we obtain from eq.(7),
$$ \vec{A}_E -\vec{A}_S \approx  $$
$$\approx \hskip3pt -
\hskip4pt \mu^*_{SE} \cdot \hskip2pt { \vec{R}_{SE} \over R_{SE}^3}
\hskip5pt  + \hskip5pt    \mu_J \Bigl [ { \vec{R}_{JS} \over R_{JS}^3}
- {\vec{R}_{JE} \over R_
{JE}^3}  \Bigl ] +
\hskip5pt \eta  \Bigl ( {\Omega \over mc^2} \Bigl )_{\hskip-2pt S}
\mu_J {\vec{R}_{JS} \over R_{JS}^3} \hskip3pt .
  \hskip5pt
\eqno(8)$$

\noindent Corresponding equations for Mars are obtained
 by replacing subscript
$E$ by $M$ in eqs.(7) and (8). To good approximation the
SEP acceleration $\vec{A}_{\eta}$
has constant magnitude and  points in the direction from Jupiter
to the Sun, and since it  depends only on the mass
distribution in the Sun, the
Earth and Mars experience the same perturbing acceleration.
The responses of
the trajectories of each of these planets  due to the   term
$\vec{A}_{\eta}$ determines the perturbation in the
 Earth-Mars range and  allows
a detection of the SEP parameter $\eta$ through a ranging experiment.

The presence of the  acceleration term $\vec{A}_{\eta}$ in the
equations of motion results in  a
polarization of the orbits of Earth and Mars,  exemplifying the
planetary SEP effect.
We investigate   here   the accuracy with
which the   parameter $\eta$ can be determined through
Earth-Mars ranging and approach the problem
with a series of successive approximations. In Sections
II and III  the ``tidal term'' $\vec{A}_{tid}$ in equation
(8) is neglected. In Section II the
perturbation theory about circular, coplanar
reference orbits for Earth and Mars is performed.
A covariance analysis is carried out to estimate the
accuracy to which the SEP parameter $\eta $ can be determined
from a lagre number of Mars ranging measurements, each of accuracy
$\sigma$ meters. In Section III   the
calculations of Section II are improved by employing
 numerical integration
rather than perturbation theory.
The agreement between the two approaches is good, and
 the eccentricity corrections are found to improve the
accuracy of the
analytic approximation significantly. In Section IV
the tidal acceleration term $\vec{A}_{tid}$,
is restored, requiring the addition of the mass of
 Jupiter $\mu_J$ to our set of covariance
parameters.  This mass   can be determined more
accurately from a few years of Earth-Mars ranging than from the
Pioneer 10,11 and Voyager 1,2 flybys combined, independent
of the $\eta$ measurement.
In Section V we summarize and suggest further avenues for
testing SEP violation.

\section{Perturbation  About a Circular Reference Orbit}

Here and in the next section  the problem is simplifyed by
ignoring the tidal
term $\vec{A}_{tid}$ in equation (8) and the corresponding equation
for Mars.  We examine the effect of the SEP acceleration
term $\vec{A}_\eta$ in (8) on the orbits of Earth
and Mars by carrying out first-order perturbation theory about the
  zeroth order orbits of Earth and Mars, taken to be circular.
Jupiter's  orbit is also taken
as circular and coplaner with Earth and Mars.
With these approximations a nine parameter covariance analysis
is carried out to estimate how precisely Mars ranging can
 determine the SEP parameter $\eta$. These   approximations
 give a standard deviation for $\eta$ which closely
agrees with the later numerical
integration result of Section III.

A heliocentric reference frame rotating with Jupiter
at constant angular frequency
$\omega = \sqrt{\mu_S/R_{SJ}^3}$ is assumed. The
SEP perturbation is represented by a
constant acceleration $g_\eta$ directed from Jupiter to the Sun.
Locating Jupiter on the $x$ axis, the following
Hamiltonian for both Earth and
Mars in polar coordinates results:
$$ H = {1 \over 2}(p_r^2 + {p_\theta^2 \over r^2}) -
\omega p_\theta - {\mu_S \over r} + g_\eta r \hskip 1mm cos \theta,
\hskip10pt  \eqno(9) $$

\noindent where
$$g_\eta \equiv \eta \biggl ({\Omega \over mc^2} \biggl)_{\hskip-3pt S}
{\mu_J \over R_{JS}^2}, \qquad
p_r = {\dot r}, \qquad \hskip10pt p_\theta =
r^2({\dot \theta} + \omega). \hskip10pt \eqno(10) $$

We take the reference orbits for Earth
and Mars to be circular Keplerian with $r = a$ and $g_\eta = 0$.
Then
$p_\theta \equiv \sqrt{\mu_S a}$ \hskip1pt and $n \equiv
\sqrt{\mu_S/a^3}$ is the orbital  angular frequency.
A covariance analysis can   be performed to
estimate the accuracy to which one can determine
the SEP parameter through Earth-Mars ranging.
 Earth-Mars range $\rho_{EM}$ is defined as follows:
$$\rho^2_{EM} = r_E^2 + r_M^2 - 2 \hskip1pt r{\hskip-1pt}_E
\hskip1pt r{\hskip-1pt}_M
 cos(\theta_M - \theta_E). \hskip10pt  \eqno(11) $$

\noindent  The  variation of $\delta \rho_{EM}$ is then:
$$\delta \rho_{EM}(t) = \sum_{k = 1}^9{\delta q_k
{\partial \rho_{EM} \over \partial q_k}(t) }, \hskip10pt   \eqno(12)$$

\noindent where $\delta \rho_{EM}$ does not depend on $\delta \omega$
and ${\bf q}$ is the following vector:
$${\bf q} \equiv  \Big(r_{E_0}, r_{M_0}, p_{Er_0}, p_{Mr_0},
p_{E \theta_0}, p_{M \theta_0},
\theta_{E_0} - \theta_{M_0}, \mu_S, g_\eta\Big).   \eqno(13) $$

\noindent  The partial derivatives are easily
computed and for example,
$$ {\partial \rho_{EM} \over \partial r_{E0}} =
{1 \over \rho_{EM}} \biggl[ a_E \hskip1pt
cos(n_E \hskip1pt t) - $$
$$- {3 \hskip1pt a_M \over 2} cos(\theta_E - \theta_M - n_E \hskip1pt t)
+ {a_M \over 2} cos(\theta_E - \theta_M +
n_E \hskip1pt t) \biggr]. \hskip10pt
 \eqno(14) $$

\noindent By definition,  the covariance matrix $\alpha_{jk}$ has elements:
$$\alpha_{jk} = \sum_{i=1}^N {{1 \over \sigma_i^2} {\partial \rho_{EM}
\over \partial
 q_j}(t_i) {\partial \rho_{EM} \over \partial q_k}(t_i) },
\hskip10pt  \eqno(15) $$

\noindent where $N$ ranging measurements have been made at
times $t_i, i=1,...,N$, and
have uncertainties $\sigma_i$.  The uncertainty
in the estimations of the SEP acceleration
term $g_\eta$ and  parameter $\eta$ are then:
$$ \sigma_{g_\eta} = \sigma_{\eta} \biggl
({\Omega \over mc^2} \biggl)_{\hskip-3pt S} {\mu_J
 \over R_{JS}^2} =
\sqrt{(\alpha^{-1})_{{g_\eta}{g_\eta}}},  \eqno(16) $$

\noindent with the fractional binding energy
 $ \big({\Omega / mc^2} \big)_S$
given by expression $(2a)$.

The results obtained are presented in Figure 1, with the thin dashed
curve showing the result of evaluating equations
(15) and (16) for $\sigma_{\eta}$
assuming $N$ daily range measurements have been
taken during the mission, each with the same
uncertainty $\sigma$, measured in meters.
The initial
angles between Earth and Jupiter and Mars and Jupiter were taken from the
DE242  emphemeris at time
2441272.75, the beginning of the Mariner 9 ranging measurements. The
quantities $n_E$ and $n_M$
are taken to be the mean motions of Earth and Mars, $a_E$ and $a_M$
their mean  distances from the Sun, and
the frequency $\omega$ is taken to be the mean motion of Jupiter.
The uncertainty in $\eta$ first
drops very rapidly with time and then after a few years approaches
the asymptotic behavior  $\sim N^{-1/2}$. This result
 gives a lower   bound on the uncertainty as
predicted  by conventional covariance analysis. For
a mission duration of   order ten years, the
uncertainty behaves as
$$\sigma_{\eta} \sim .0028 \sigma/\sqrt{N}   \eqno(17)$$

This result,  eq.(17), assumes Gaussian random ranging errors
with a white spectral frequency distribution. But
past ranging measurements using the Viking Lander have
been dominated by systematic error (Chandler et al. 1994).
One approach to accounting for systematic error is to
multiply  the formal errors from the   covariance matrix
by $\sqrt{N}$ (Nordtvedt  1978).  With this approach,
the expected error decreases rapidly near the beginning of the
data interval, but for large $N$ approaches an asymptotic value
as demonstrated by Fig.1.  However, we believe this is overly
conservative.  A more optimistic error estimate would include
a realistic description of the time history
of the systematic error.  For example if we
knew that the systematic error was a sinusoid of known angular
frequency, we could add its amplitude and phase as additional
variables in the covariance analysis.  By contrast the worst-case
error analysis treats each variable of concern equally by
multiplying its computed error by $\sqrt{N}$.

A realistic systematic error budget for ranging
data to Mars, or for   Mercury as considered by
a group at University of Colorado (Ashby
et al. 1995), is not presently available.
But it is unlikely
that we will be so unfortunate that the
frequency spectrum of the
signal will correspond to the spectrum of the systematic error.
However, we reduce the upper error bound determined by
the $\sqrt{N}$ multiplier ($\sigma_\eta = 0.0028 \sigma$)
 by a numerical factor.  The
ranging experiment proposed by Ashby et al. (1995) for Mercury is
quite similar to our proposed experiment using Mars.
We therefore
follow the Colorado group and reduce the worst-case
error estimate by a factor of three and call the result the
modified worst-case analysis.  This yields an asymptotic value for
the error of $\sigma_\eta = 0.0009\sigma$, in our opinion
a realistic estimate of the upper bound on the error.

In the case of the existing Mars ranging derived from the
Mariner 9, Viking, and Phobos missions, the rms ranging residual
referenced to the best-fit Martian orbit was 7.9 m.  We computed
the covariance matrix with assumed daily range measurements for
Mariner 9 (actual data interval JD 2441272.750 to JD 2441602.504)
and Viking (actual data interval JD 2442980.833 to JD
2445286.574).  Additionally,  one ranging measurement
from Phobos (actual time JD 2447605.500) was included, although it had
negligible effect on the result.  With $\sigma$ = 7.9 m,
a formal error $\sigma_\eta = 0.0005$ is obtained from the
covariance matrix. If 7.2 years of
Mars ranging is   assumed, though not continuous, we reach the
asymptotic limit of the modified worst-case analysis
 (as shown by Fig. 1), a
realistic error $\sigma_\eta = 0.009$ which is about 17 times the
formal error.  This is a factor of eight larger than
the realistic error set by Chandler et al. (1994)
from an analysis
of the actual combined LLR and Mars ranging data.
We conclude that the
best determination of $\eta$ is provided by the LLR data, but the
existing Mars ranging can provide an independent solar test
with  a realistic accuracy interval of
$$  \sigma_{\eta}\approx  0.0005 -  0.009 \hskip15pt
({\rm Mariner \hskip 5pt 9,\hskip 5pt Viking,
\hskip 5pt Phobos}). \hskip10pt \eqno(18) $$

  Expression (35) establishes the   interval for
expected accuracy $\sigma_{\eta}$ with
the lower and upper bounds  estimated for existed
data by conventional covariance
 analysis and   ``modified worst case'' analysis  respectively.
 This   interval will be narrowed by ongoing upgrades to DSN
instrumentation and better modelling of the
antenna and spacecraft ranging systems
(Anderson et al. 1985). Future Mars
Orbiter and Lander missions are expected to achieve  an rms
systematic ranging error between 0.5 and 1.0 m. Then   after  a few
years of ranging, the realistic error  on $\eta$ should fall to
around
$$  \sigma_{\eta}\approx  \left(0.00006 -
0.0011\right)\sigma \hskip15pt
({\rm Future \hskip 5pt Mars \hskip 5pt
 missions }). \hskip10pt \eqno(19) $$

\section{Numerical Integration  Without the
 Tidal Term $\vec{A}_{tid}$.}
To obtain a more accurate estimation
of $\sigma_{\eta}$ and to check
 the results of Section II,   a
numerical integration of the Sun-Earth-Mars-Jupiter system in
heliocentric coordinates is performed.  Equation (8) without
 the $\vec{A}_{tid}$ term was used for Earth and Mars
(with the interchanging of the subscripts
$(E) \rightarrow  (M)).$
The equation of motion for Jupiter did not include a SEP term
or Newtonian perturbation from other planets.
The DE242 ephemeris was again used for the initial conditions, and
the same 9 parameters of equation (30) were used in the
covariance analysis.

Assuming frequent  Earth-Mars range measurements as in Section II,
the  results are
shown in Figure 1 (thick dashed curve).
The results agree fairly well with the analytical ones of
Section II (thin dashed curve).  For  ten years
of observations  we obtain
$$\sigma_{\eta} \sim .0027 \sigma/\sqrt{N}, \eqno(20)$$

\noindent with range measurement $\sigma$ in meters.
  Note that this result compares favorably with the cruder
analytic result presented by expression (17).

An estimate of how well $\eta$
could be estimated from existing ranging
data, as discussed at the end of Section II, yields:
$$ \sigma_{\eta} \approx 0.00055 - 0.009 \hskip15pt
({\rm Mariner \hskip 5pt 9,
\hskip 5pt Viking, \hskip 5pt Phobos}), \hskip10pt  \eqno(21) $$

\noindent which is slightly smaller than the
analytic result given in eq.(18).

While numerical integration is expected to be
more accurate than   analytic
approximations, one might  wish to gain more understanding
of the planetary SEP effect by doing a   realistic analytic
calculation which improves on  the
"circular orbit" approximation of Section II.
It is natural to
eliminate the largest sources of error of that approximation.
Mars has an orbital eccentricity of $.093$,  and
there is no fundamental barrier
to using elliptical reference orbits. The analytic
 calculation of Section II was redone with elliptical
reference orbits for Earth and Mars, working to first order
in the eccentricity.  We also used  a more sensible way
to include the eccentricity corrections, the method of
variation of parameters (Robertson \& Noonan 1968), and were able
to solve the variation of parameters
equations for the perturbed orbits of Earth and Mars
to fourth order in the eccentricity.
Figure 2 shows  plots of the partial
derivative of the Earth-Mars range with respect to the
SEP acceleration.
Comparison of the three curves shows that the eccentricity
correction plays a more fairly significant role,
  than one might expect.
One reason for this is that the eccentricity
 corrections
turn out to include more ``secular'' matrix elements which are
 proportional to the time $t$.
Such  elements  dominate at large times, and
  the eccentricity corrections thereby qualitatively change
the nature of the solution in the linear approximation.

\section{Numerical Integration With the
 Tidal Term $\vec{A}_{tid}$.}

  Jupiter's mass $\mu_J$ needs treatment
as  an adjastable parameter
to be  fit with the ranging data. This is because
the octopolar tide of Jupiter acting on the orbits of
Earth and Mars produces polarizations similar to
those produced by the SEP effect, but fortunately having a
different Earth-Mars ratio
 and   therefore   separable from the
desired SEP effect. If Jupiter's mass were uncertain by
 4 parts in $10^8$, its tidal polarization
of Mars' orbit would be uncertain by that
would be produced by an $\eta \sim .001$, for example.
But Jupiter's mass is only
 known to a part in a million, so we must include
$\mu_J$ as a free parameter with its own partial.

In this Section we outline the most accurate calculation.
The full equations  (8) (and analogue for Mars) were
numerically integrated including
the tidal term $\vec{A}_{tid}$, previously was
neglected in Sections II and III.
To the parameters
$r_{E_0}, r_{M_0}, p_{Er_0}, p_{Mr_0}, p_{E \theta_0},
p_{M \theta_0},$   $\theta_{E_0} -
 \theta_{M_0}, \mu_S$ and $g_\eta$,then, we add $\mu_J$
in the covariance analysis, otherwise, the analysis is
identical to that of Section III.

The results are shown in Figure 1, with the solid curve
giving the result for
$\sigma_\eta$ from $N$   ranging measurements, each
with error $\sigma$ meters.
For a mission time of the order ten years we find
$$\sigma_{\eta} \sim .0039 \sigma/\sqrt{N}.   \eqno(22)$$

The estimation of how well $\eta$ can be determined from existing
ranging data (as discussed in Section II) is:
$$ \sigma_{\eta} \approx 0.0012- 0.02 \hskip15pt
 ({\rm Mariner \hskip 5pt 9,
\hskip 5pt Viking, \hskip 5pt  Phobos} ), \hskip10pt
 \eqno(23) $$

\noindent about double the result from Section III.

The covariance analysis gives the expected
formal error in $\mu_J$ as well, analogous to eq.(16),
with the  result    shown in Fig. 3.  For a mission time
of order ten  years
we find $\sigma_{\mu_J} \sim 5.7  \sigma/\sqrt{N}$
 in $km^3s^{-2}$, where $N$,
as before, is the number of daily ranging measurements taken
 during the mission.  For
 $\sigma = 7.9 m$, $\sigma_{\mu_J}$ falls below
the present accuracy
 determined from Pioneer 10,11 and Voyager
1,2, namely, $\mu_J = 100 km^3/s^2$ (Campbell \& Synnott 1985) -
within two years. This surprising result suggests that
Earth-Mars ranging can provide an extremely accurate
value for the mass of Jupiter, independent of the
any determination   of a SEP effect.

Following   Section II, in order to obtain the
 realistic estimates for $\sigma_{\eta}$ and $\sigma_{\mu_J}$
 one   multiplyes
these results by $\sqrt{N}/3$,   giving values of the
upper bound of realistic errors for $\eta$ and $\mu_J$ as
$\sigma_{\eta} \sim .0012 \sigma$ and $\sigma_{\mu_J}
\sim 1.9  \sigma$  $km^3s^{-2}$ correspondingly.
With mission duration of order ten
  years  this result gives
the interval for the uncertainties $\sigma_{\eta}$
and $\sigma_{\mu_J}$
during  the future Mars missions:
$$ \sigma_{\eta} \approx \left(0.0001 - 0.0012\right)\sigma, $$
$$ \sigma_{\mu_J}\approx \left(0.09 - 1.9\right)\sigma
\hskip 10 pt {\rm km^3/s^2}.\eqno(24)$$

\section{Discussion}

 The  planet Mars has become an object of intensive
investigation by many scientists around
 the world. The next flight opportunity during 1996-97
will mark the initiation of a number of new  space
 missions to that planet
from which we expect to obtain a rich set of
data, including spacecraft tracking and planetary radar
measurements, and allowing precise relativistic
celestial mechanics experiments.

Anticipating these events, we have analyzed the ability for
testing  SEP violation with
Earth-Mars ranging. The expected accuracy of the
future ranging
experiments would put significant constraints on
theoretical models, including a possible
inequality of the Sun's inertial and gravitational masses.
Using analytic
 and numerical methods we have shown from  covariance analysis
that  Earth-Mars ranging can provide a quality  estimate of
$\eta$.
Indeed, for $N$ ranging measurements  with an accuracy of
 $\sigma$ meters, the SEP parameter
 $\eta$ according to covariance analysis  can be
determined within the  accuracy
$\sigma_\eta \hskip 2pt \approx \hskip 2pt 3.9 \times 10^{-3}
\hskip 2pt {\sigma /\sqrt{N}}.$
The ``realistic'' estimate for $\sigma_\eta$ based on a
``modified worse case'' analysis sets a conservative
limit on the accuracy and  indicates that
even in   unfavorable cases
the Sun-Earth-Mars-Jupiter system allows for a sensitive test
of the Strong Equivalence Principle,
qualitatively different from that provided by LLR.
The mass of Jupiter, $\mu_J$, can be determined more
accurately from a few years of Earth-Mars ranging than from
Pioneer 10,11 and Voyager 1,2 combined. This analysis
shows a rich opportunity for  obtaining new scientific
results from the the program of ranging measurements to Mars.

Efforts are underway at JPL to determine $\eta$ from the
Mariner 9, Viking and Phobos ranging data. This
research   will modify the  theoretical model to
include effects due to Saturn. We will perform
the numerical experiments with combined data collected
from the planetary missions, LLR and VLBI.  And, as the
data do not presently include any direct ranging to the Sun, it will
 be interesting to include
ranging results to the spacecrafts of the joint US-Russian
Solar probe missions scheduled for launch in the year 2001,
or shortly thereafter.
A preliminary determination from combined solar-system data,
 including Mars ranging and lunar-laser
ranging, has been reported at a Division of Dynamical
Astronomy meeting by Chandler et al. (1994).
However, we have found that the inclusion of the
 $\vec{A}_\eta$ acceleration of eq.(7)  in the JPL
planetary ephemerides system as postulated
(Standish  et al. 1993), has not been
straightforward.
The total SEP range signal in Earth-Mars ranging is so
complex and unique, one should be
very conservative about physical interpretation of the
 results obtained. The full scale research of this
important experiment is currently underway at JPL.
The results reported here provide insights
into what is being measured, and hence they minimize the
possibility of error in implementing the planetary SEP
effect in  the complicated software system.
And by concentrating on covariance analysis,
information needed for the planning of
gravitational experiments on future Mars missions is
 obtained. We intend that this paper serve as one of a collection
of quides for scientific goals and priorities on
Mars missions over the next decades.

Finally we mention that the analysis of solar ranging data
might provide the opportunity for another fundamental test, namely a
Solar system search for dark matter (Nordtvedt  1994;
Braginsky  1994; Nordtvedt 1995). Suppose that dark matter  weakly
interacts  with ordinary matter in a manner depending
on the specific properties of the matter composing the bodies.
The  Sun, having an internal
structure and matter composition which is considerably different
from the rest of the inner bodies in the
Solar system, might then have a
different coupling to dark matter, and a
corresponding anomalous cosmic force $\vec{F_c} = m_S\vec{A}_c$
acting on the Sun, would produce extra terms in the heliocentric
 equations of motion for the planets - like
$\vec{A}_\eta$, but fixed in direction, in equation (8).
Interesting limits
on the size of any SEP violation for extended
bodies in the Solar system falling toward dark matter may be obtainable.
This research will be the subject of a subsequent publication.

We are indebted to our colleagues John Armstrong, Eunice Lau
and Skip Newhall  for many useful and
stimulating conversations. MG acknowledges the partial support
 by an AWU-JPL sabbatical fellowship.
KLN was supported in part by National Aeronautic and Space
Administration  throughout Contract NASW-4840.
SGT was supported by  National Research Council,
Resident Research Associateship award.
This work was carried out in part at the Jet Propulsion
Laboratory, California Institute of Technology,
under a contract with National Aeronautic and Space
Administration.

\section{References}

\noindent Adelberger E. G. et al. 1994, Phys. Rev., D50, 3614

\noindent  Allen, C. W.  1985,  Astrophys. Quantities
 (London: Athlone press)

\noindent Anderson, J. D., Levy, G. S. \& Renzetti, N. A.
1986, in:  Realativity in Celestial Mechanics and
Astrometry  (IAU Simposium 114), eds. J. Kovalevsky
\& V. A. Brumberg (Reidel: Dorthecht), 329

\noindent Ashby, N., Bender, P. L. \&  Wahr, J. M.
1995, private communication

\noindent Braginsky, V. B. \& Panov, V. I. 1972,
 Sov. Phys. JETP,  34, 463

\noindent Braginsky, V. B. 1994, Class. Quantum Grav., 11, A1

\noindent Bertotti, B. \& Grishchuk L. P. 1990,
 Class. Quantum Grav.,  7, 1733

\noindent Campbell, J. K. \& Synnott, S. P. 1985, AJ,   90 , 364

\noindent Chandler, J. F.,  Reasenberg, R. D. \&
Shapiro I. I.  1994, BAAS, Vol. 26, No.2, 1019

\noindent Damour, T. \& Sch\"afer G. 1991,  Phys. Rev. Lett.,
 66,  2549

\noindent Dickey, J. O., Newhall, X X \& Williams, J. G.  1989,
Advance in Space Research, 9 , 75

\noindent Dickey, J. O.   et  al.  1994, Science,  265 , 482

\noindent Orellana, R. B  \& Vucetich H.  1993, A\&A, 273, 313.

\noindent Nordtvedt, K., Jr.  1968a, Phys. Rev.,   169 , 1014

\noindent $\underline{\hskip 40pt}$. 1968b, Phys. Rev.,   169, 1017

\noindent $\underline{\hskip 40pt}$. 1968c, Phys. Rev.,   170 , 1186

\noindent $\underline{\hskip 40pt}$. 1970,  Icarus,   12 , 91

\noindent $\underline{\hskip 40pt}$. 1973,  Phys. Rev., D 7, 2347

\noindent $\underline{\hskip 40pt}$. 1978,  in "A Close-Up of the Sun"
 (JPL Pub. 78-70), 58

\noindent $\underline{\hskip 40pt}$. 1994,  ApJ,   437, 529

\noindent $\underline{\hskip 40pt}$. 1995,  Icarus,   114, 51

\noindent Robertson, H. P. \& Noonan, T. W.  1968,
  Relativity and Cosmology, (Philadelphia: Saunders)

\noindent Roll, P. G., Krotkov, R. \& Dicke, R. H.  1964,
Ann. Phys. (N.Y.),   26, 442

\noindent Singe, J. L. 1960,   Relativity: the General Theory
(Amsterdam: North-Holland)

\noindent Shapiro, I. A.   et al. 1976, Phys. Rev. Lett.,   36, 555

\noindent Standish, E. M.  et al.  1992, in:  Exp. Suppl.
to the Astron. Almanac,  (Mill Valley: Univ. Sci. Books), 279

\noindent Ulrich, R. K.  1982, ApJ, 258, 404

\noindent Will, C. M. \& Nordtvedt, K.  1972, ApJ, 177, 757

\noindent Will, C.M.  1971, ApJ, 169, 125

\noindent $\underline{\hskip 40pt}$. 1993,
Theory and Experiment in Gravitational Physics,
(Cambridge: Cambridge  Univ. Press)

\noindent Williams, J. G.   et al.  1976, Phys. Rev. Lett., 36, 551

\noindent Williams, J. G., Newhall X X \& Dickey, J. O.  1995,
  Submitted to  Phys. Rev., D

\section{Figure Legends}

{\bf Fig. 1}.  Variation of the uncertainty in the
SEP parameter $\eta$ with $N$,
 assuming $N$  Earth-Mars
range measurements, each with uncertainty $\sigma$ meters.
The thin dashed curve
 is the result of
the analytic approximations of Section II.
The thick dashed curve is the
numerical integration result of Section III.
The solid curve comes from the numerical integration
described in Section IV
where the term $\vec{A}_{tid}$ was restored
and $\mu_J$ was included in the covariance analysis.

\vskip 10pt
{\bf Fig. 2}.  Investigation of first order eccentricity
corrections (thick dashed
curve) to the linear approximation of Section II
(thin dashed curve).
Both of these curves should be compared to the
more accurate numerical results
of Section III (solid curve).
All three of these calculations used
eq.(8) without the $\vec{A}_{tid}$ term.  The
initial conditions taken were
different from those in figures 1 and 3.

\vskip 10pt
{\bf Fig. 3}.  Plot of the uncertainty in the mass of
Jupiter versus $N$ from the numerical
integration of equation (8) plus covariance analysis.
See Section IV.
\vskip 10mm

\end{document}